\documentclass[%
 reprint,
superscriptaddress,
 amsmath,amssymb,
 aps,
]{revtex4-1}

\usepackage{graphicx}
\usepackage{dcolumn}
\usepackage{bm}
\usepackage{xcolor}
\usepackage{hyperref}

\begin{document}

\title{Strain-engineering of Berry curvature dipole and \\valley magnetization in monolayer MoS$_2$}

\author{Joolee Son}
 \affiliation{Department of Physics and Department of Energy Systems Research, Ajou University, Suwon, 16499, Korea}
 
 \author{Kyung-Han Kim}
 \affiliation{Department of Physics, Pohang University of Science and Technology, Pohang, 37673, Korea}
 
 \author{Y. H. Ahn}
 \affiliation{Department of Physics and Department of Energy Systems Research, Ajou University, Suwon, 16499, Korea}
 
 \author{Hyun-Woo Lee}
  \email{hwl@postech.ac.kr}
 \affiliation{Department of Physics, Pohang University of Science and Technology, Pohang, 37673, Korea}
 
 \author{Jieun Lee}
  \email{jelee@ajou.ac.kr}
 \affiliation{Department of Physics and Department of Energy Systems Research, Ajou University, Suwon, 16499, Korea}


\begin{abstract}
The Berry curvature dipole is a physical quantity that is expected to allow various quantum geometrical phenomena in a range of solid-state systems. Monolayer transition metal dichalcogenides provide an exceptional platform to modulate and investigate the Berry curvature dipole through strain. Here we theoretically demonstrate and experimentally verify for monolayer MoS$_2$ the generation of valley orbital magnetization as a response to an in-plane electric field due to the Berry curvature dipole. The measured valley orbital magnetization shows excellent agreement with the calculated Berry curvature dipole which can be controlled by the magnitude and direction of strain. Our results show that the Berry curvature dipole acts as an effective magnetic field in current-carrying systems, providing a novel route to generate magnetization.
\end{abstract}

\maketitle


Berry curvature is central to various topological phenomena observed in solid-state crystals \cite{Xiao2010}, ultracold atoms \cite{Tarruell2012,Aidelsburger2013} and photonic architectures \cite{Lu2014}. In view of charge transport, its effect has been assumed to be limited to magnetic systems with broken time reversal symmetry as hallmarked by the anomalous Hall effect \cite{Nagaosa2010}. Recent theories, however, demonstrated that in nonlinear regime, Hall effect can occur even in time-reversal symmetric systems if they are noncentrosymmetric and possess reduced symmetry (\emph{e.g.}, only one mirror plane) \cite{Sodemann2015,Low2015}. The nonlinear Hall effect has been attributed to the dipole moment formation of the Berry curvature in momentum space. Such Berry curvature dipole widens the scope of Berry curvature effects. It was recently proposed that in transition metal dichalcogenides (TMDs), the Berry curvature dipole may be generated when spatial symmetries of TMDs are lowered by strain \cite{Sodemann2015,You2018}. This motivates monolayer TMDs as promising materials venue to examine the Berry curvature dipole. As exemplified by ultracold gases, where the in-depth examination of the Berry curvature effects becomes possible through the Berry curvature variation by the optical lattice modulation \cite{Tarruell2012,Aidelsburger2013}, the mechanical tunability of the Berry curvatures in TMDs provides an ideal avenue to explore the Berry curvature dipole. 

\begin{figure}[t]
    \centering
    \includegraphics[width=0.5\textwidth]{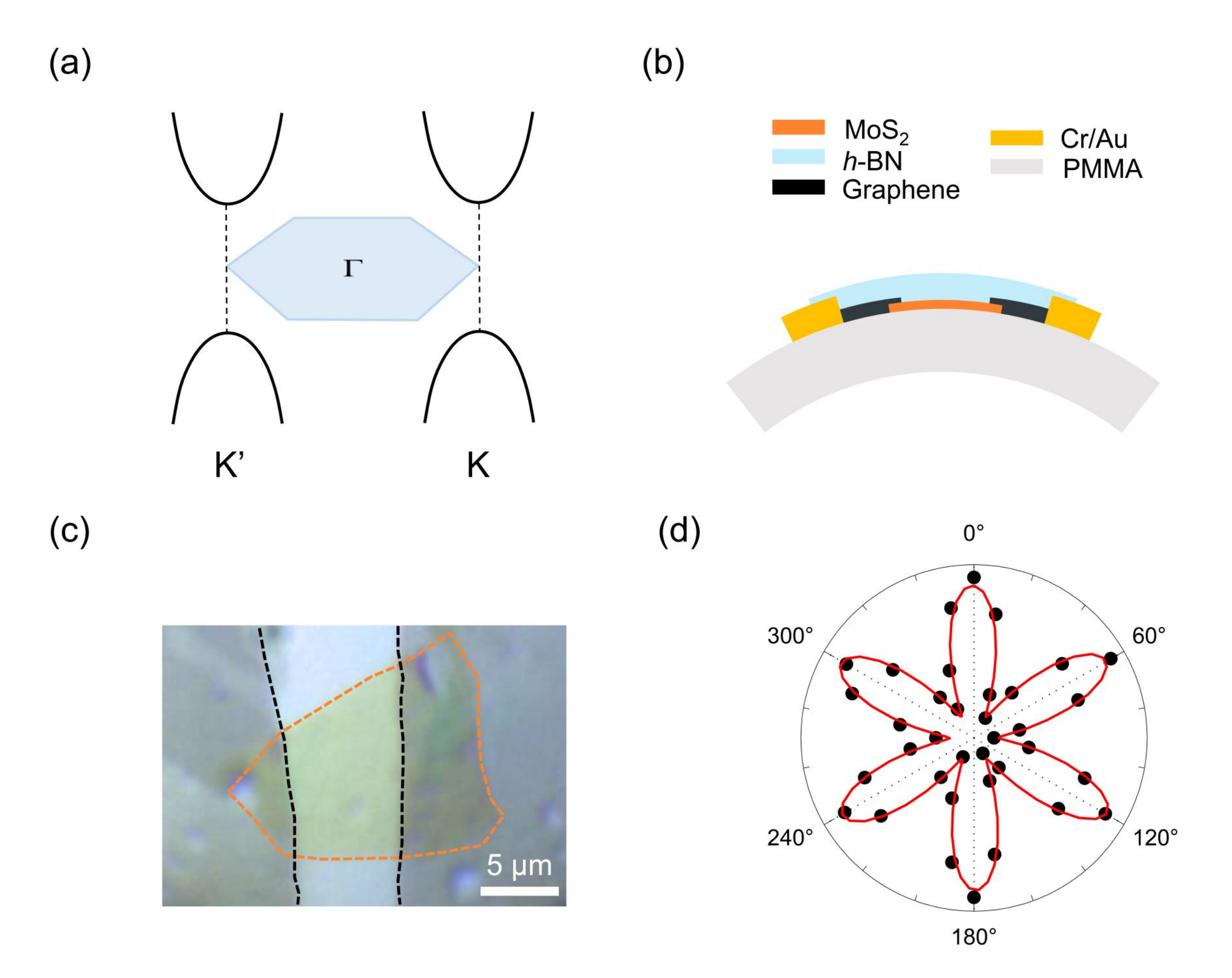}
    \caption{Strain-tunable monolayer MoS$_2$. {\bf (a)} $K$ and $K^\prime$ valleys of monolayer MoS$_2$ in momentum space. {\bf (b)} Device schematics of van der Waals heterostructure fabricated on flexible polymethyl methacrylate (PMMA) substrate with pre-patterned Cr/Au metal pads. {\bf (c)} Microscope image of a device consist of MoS$_2$ (orange dashed line), graphene (black dashed line) and capping {\it h}-BN. {\bf (d)} SHG measurement for identifying the crystal orientation of MoS$_2$ shown in (c). The angles of maximum SH intensity repeating every 60$^\circ$ correspond to the armchair direction of the monolayer crystal.
    }
    \label{Fig1}
\end{figure}

TMDs are noncentrosymmetric two dimensional materials and have two nonequivalent $K$ and $K^\prime$ valleys holding the opposite signs of the Berry curvature \cite{Xiao2012} (Fig. \ref{Fig1}(a)). Such valley-dependent Berry curvature gives rise to intriguing phenomena \cite{Xu2014,Mak2018} such as the valley optical selection rule \cite{Cao2012,Mak2012,Zeng2012}, the valley Zeeman effect \cite{Li2014,MacNeill2015,Srivastava2015,Aivazian2015} and the valley Hall effect \cite{Mak2014,Lee2016,Lee2017}. Further manipulation of electrically induced valley magnetization has been demonstrated by lowering the crystal symmetries of TMDs by strain \cite{Lee2017}, but the dependence on strain yet remains unclear. Given the inherent flexibility of monolayer TMDs, comprehensive studies on strain dependence may allow simultaneous investigation of the valley magnetization and Berry curvature dipole. In this work, we employ stretchable van der Waals heterostructures and demonstrate that the strain-induced Berry curvature dipole in monolayer molybdenum disulfide (MoS$_2$) generates electrically induced valley orbital magnetization. The magnitude and direction of the valley orbital magnetization are directly probed by Kerr rotation microscopy as a function of strain, which shows an excellent agreement with calculated Berry curvature dipole. The observed valley magnetization increases with the increasing magnitude of strain and \emph{reversibly} turns on and off and flips the sign depending on the direction of strain. Our approach demonstrates strain engineering as a distinct pathway to control the valley degree of freedom in monolayer TMDs and provides a precision tool to map out the Berry curvature dipole in solid-state crystals.

The van der Waals heterostructures in our work consist of monolayer MoS$_2$, few-layer (FL) graphene and hexagonal boron nitride ({\it h}-BN) as shown in Fig. \ref{Fig1}(b). The exfoliated monolayer MoS$_2$ crystal is first transferred onto a flexible substrate with pre-patterned metal pads. Then two FL graphene exfoliated onto SiO$_2$/Si substrate are picked up by {\it h}-BN using poly-propylene carbonate (PPC). The {\it h}-BN and two graphene stack is transferred onto MoS$_2$ to form source and drain electrodes. The {\it h}-BN capping layer, with a thickness around 20 nm unless otherwise noted, minimized the exposure of MoS$_2$ to air and enhanced the adhesion between MoS$_2$ and substrate. The final device image is shown in Fig. \ref{Fig1}(c). We note that all our measurements in this work are performed at room temperature.

To apply strain in desired directions, the crystal orientation of the MoS$_2$ is identified by the polarization-resolved second harmonic generation (SHG) \cite{Li2013,Malard2013}. In this measurement, the SH intensity component parallel to the polarization of the pump laser is detected. Fig. \ref{Fig1}(d) shows the polar plot of the SH intensity measured as a function of the laser polarization angle, which shows the maximum intensity when the polarization aligns to the armchair direction of the crystal (Methods in Supplementary Materials). 

\begin{figure}[t]
    \centering
    \includegraphics[width=0.5\textwidth]{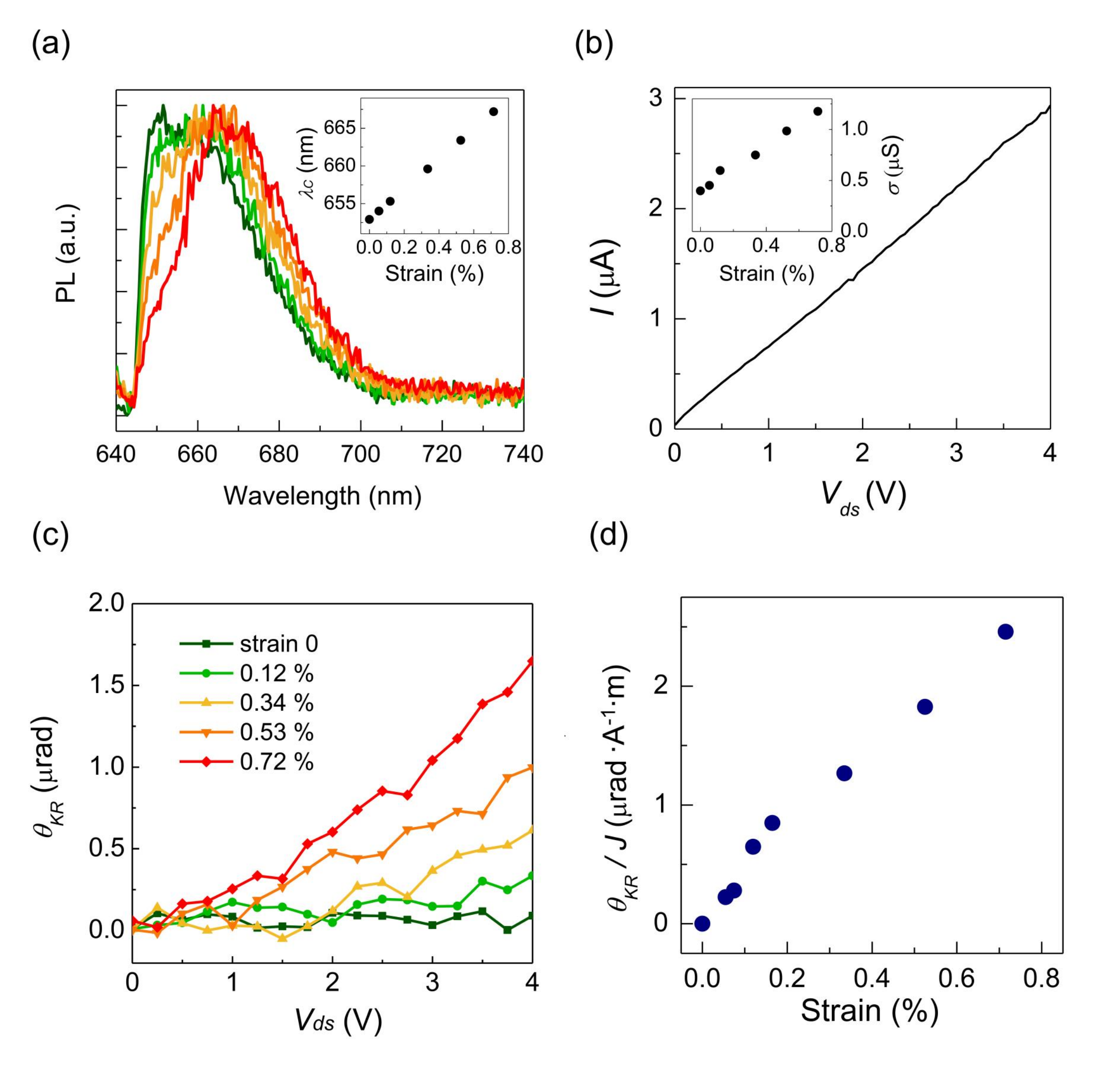}
    \caption{Valley magnetization measurement on a monolayer MoS$_2$. {\bf (a)} PL with increasing strain. Inset: Strain dependence of the PL center wavelength ($\lambda_c$). {\bf (b)} $I-V$ curve of the MoS$_2$ channel at zero strain. Inset: Channel conductance as a function of strain. {\bf (c)} $\theta_{\rm KR}$ measured at the center of the MoS$_2$ channel as a function of bias voltage ($V_{\rm ds}$) with increasing strain. {\bf (d)} Strain dependence of $\theta_{\rm KR}$ normalized by the channel current density ($J$).
    }
    \label{Fig2}
\end{figure}

With information on the crystal orientation, we applied strain in the armchair direction of the crystal by bending the substrate. The magnitude of strain is obtained from the bent geometry of the substrate and further confirmed by photoluminescence (PL) measurement of MoS$_2$ (Fig. \ref{Fig2}(a)). The PL red-shifts of 20 nm per strain percent is repeatedly observed from several different devices, showing the accurate determination of strain level in our experiment (Supplementary Materials Fig. S1). The observed shift ratio reflects the strain-induced modification of the bandgap which is independent of the strain direction \cite{He2013}. The $I-V$ curve is monitored on the same device as shown in Fig. \ref{Fig2}(b). The measured conductance is found to increase with strain because of the increasing contact area between graphene and MoS$_2$ caused by bending.

While applying strain on the device, the magnetization of MoS$_2$ is simultaneously measured at the channel center as functions of strain and an in-plane bias. The magnetization is directly probed by using the Kerr rotation (KR) spectroscopy. For KR measurement, a linearly polarized laser with a wavelength tuned to the transition resonance of MoS$_2$ is focused onto the sample and polarization rotation ($\theta_{\rm KR}$) of the probe laser upon reflection is measured by balanced detectors (Methods in Supplementary Materials). To increase the signal-to-noise ratio, we measured $\theta_{\rm KR}$ that is frequency-locked to the alternating bias applied to the source electrode. When there is no strain applied to MoS$_2$, negligible $\theta_{\rm KR}$ is measured at a finite bias. However, by applying strain, nonzero $\theta_{\rm KR}$ is detected which increases with strain and an in-plane bias (Fig. \ref{Fig2}(c)). To examine the dependence of $\theta_{\rm KR}$ on strain under the same current level, $\theta_{\rm KR}$ normalized by the channel current density ($J$) is plotted as a function of strain in Fig. \ref{Fig2}(d). The measured  $\theta_{\rm KR}$/$J$ shows linear dependence on the magnitude of strain. Similar strain and bias dependence of $\theta_{\rm KR}$/$J$ is measured from several different devices as exemplified in Supplementary Materials Fig. S2.

\begin{figure}[t]
    \centering
    \includegraphics[width=0.5\textwidth]{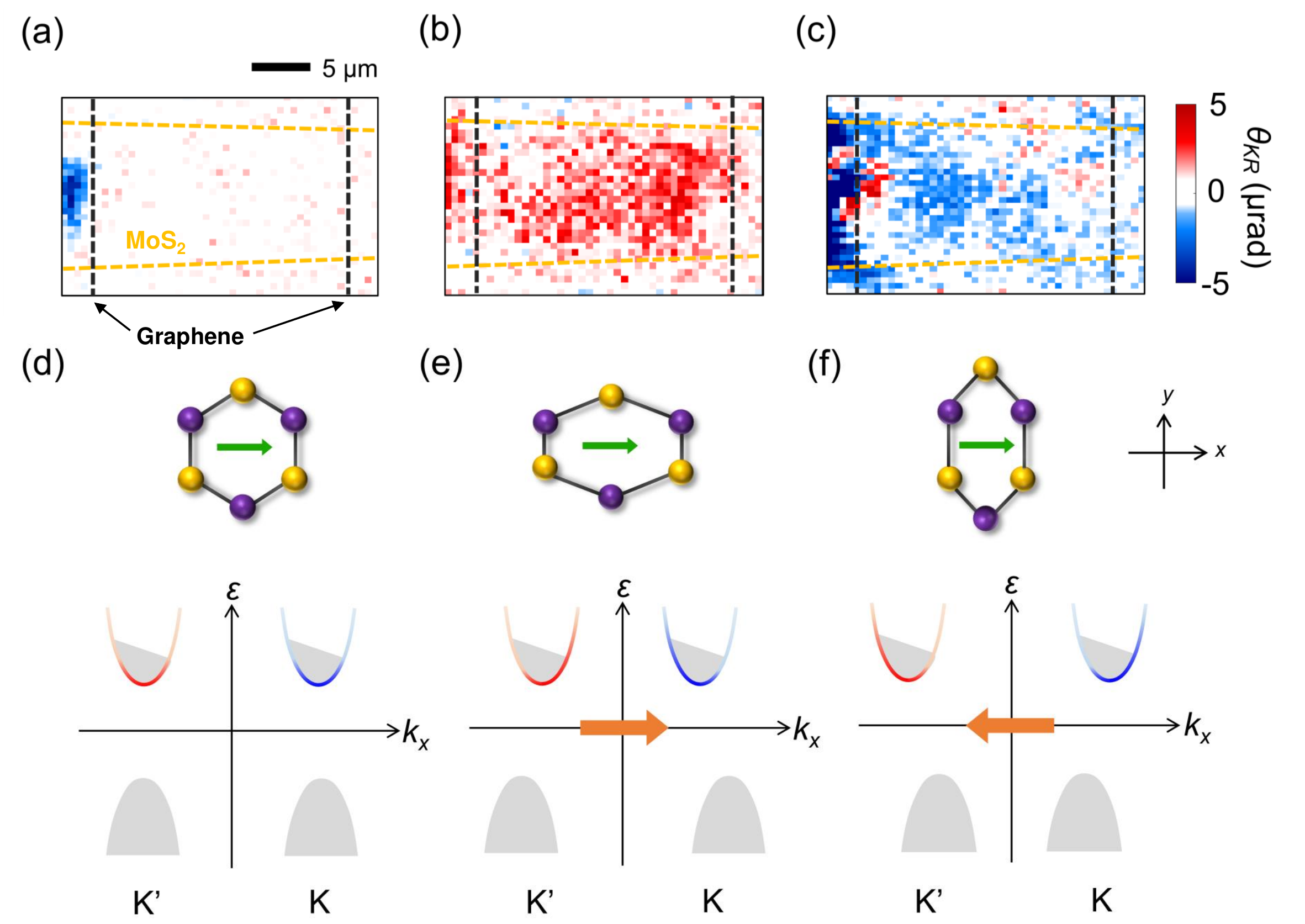}
    \caption{Strain direction dependence of the valley magnetization. {\bf (a)} KR map measured under zero strain at a finite bias ($V_{\rm ds}=3.5$ V). The boundaries of MoS$_2$ and graphene are shown by yellow and gray dashed lines, respectively. {\bf (b)} KR map when 0.55 \% strain is applied along $\hat{x}_{\rm zigzag}$. {\bf (c)} KR map when 0.55 \% strain is applied along $\hat{y}_{\rm armchair}$. {\bf (d-f)} Upper: Top-view of the MoS$_2$ crystal structure under variable strain. Purple, Molybdenum; Yellow, Sulfur. Lower: Schematics of modified band structure and Berry curvature distribution under variable strain. The red and blue colors represent positive and negative signs of the Berry curvature, respectively. The polarization of the Berry curvature distribution produces Berry curvature dipole (${\bf D}$) (orange arrows). The application of an in-plane electric field (${\bf E}$) (green arrows) generates Fermi level tilting in the conduction band.
    }
    \label{Fig3}
\end{figure}

We then investigated the strain direction dependence of the magnetization as shown in Fig. \ref{Fig3}. The armchair direction of monolayer MoS$_2$ in this device is aligned parallel to graphene electrodes so that the channel current flows in the zigzag direction ($\hat{x}_{\rm zigzag}$) of the crystal, and tensile strain is applied in one of zigzag ($\hat{x}_{\rm zigzag}$) or armchair ($\hat{y}_{\rm armchair}$) direction. For this measurement, $\theta_{\rm KR}$ on the entire channel is imaged as a function of position by scanning the sample position. Fig. \ref{Fig3}(a) shows null KR map of the device when there is no strain under a finite bias voltage ($V_{ds}$ = 3.5 V). Under the same bias, by applying strain along $\hat{x}_{\rm zigzag}$ as shown in Fig. \ref{Fig3}(b), a finite $\theta_{\rm KR}$ with positive sign is measured on the entire channel. From the 2D scan map of $\theta_{\rm KR}$ that was detected primarily on MoS$_2$ channel area, we confirm that the measured signal originates from the magnetization of MoS$_2$. Also, the measured values of Kerr rotation as well as Kerr ellipticity showed negligible dependence on the probe polarization angle, ruling out the possibility that the strain- or current-induced birefringence is detected (Supplementary Materials Fig. S3). Then we released the strain and applied the same amount of strain again along $\hat{y}_{\rm armchair}$ (PL data in Supplementary Materials Fig. S4). Surprisingly, the sign of $\theta_{\rm KR}$ changes all over the channel under the armchair strain (Fig. \ref{Fig3}(c)). The direction and magnitude of the channel current were maintained to be the same under two different strain directions. Below we show that the observed magnetization in monolayer MoS$_2$ that is switchable with strain at room temperature is possible due to the mechanical control of the Berry curvature dipole.

To verify the origin of the observed magnetization, we performed a tight binding model calculation and obtained energy band and Berry curvature distribution of monolayer MoS$_2$ at different strains \cite{Neto2009,Vozmediano2010,Peng2013,Gomez2013,Zhang2013,Rostami2015}. Details can be found in Supplementary Materials Sec. 3.1. In short, the application of strain modifies the Berry curvature (${\bf \Omega}$) distribution about $K$ and $K^\prime$ points (red and blue colors of conduction band). Unlike pristine monolayer MoS$_2$, the application of strain shifts the band edge in the opposite directions depending on the strain orientation as shown in Fig. \ref{Fig3}(e) and (f). Note that ${\bf \Omega}$ becomes asymmetric with respect to the bottom of the conduction band, implying the emergence of the Berry curvature dipole (${\bf D}$) (marked by orange arrows), where $D_a = \int \frac{d^2 {\bf k}}{(2\pi)^2} f({\bf k}) \partial_{k_a} \Omega_z({\bf k})$ [6]. Here, ${\bf k}$ is the crystal momentum, $f({\bf k})$ is the electron occupation function and $a=x, y$. For pristine monolayer MoS$_2$ with 3-fold rotational symmetry, ${\bf D = 0}$ (Fig. \ref{Fig3}(d)). Strained monolayer MoS$_2$, on the other hand, possesses asymmetric ${\bf \Omega}({\bf k})$ leading to the nonvanishing ${\bf D}$ which is perpendicular to the mirror plane and thus aligned along the crystal's mirror line, {\it i.e.}, $\hat{x}_{\rm zigzag}$. More interestingly, the direction of ${\bf D}$ is opposite for two different strain directions; along the zigzag (Fig. \ref{Fig3}(e)) and armchair (Fig. \ref{Fig3}(f)).

We now demonstrate that strain-induced Berry curvature dipole becomes the source of magnetization when there is an in-plane electric field. Consider the orbital magnetization (${\bf M}$) defined as \cite{Xiao2005,Yoda2015}
\begin{equation}
{\bf M} = \frac{1}{t} \int \frac{d {\bf k}}{(2\pi)^2} f({\bf k}) \left\{ {\bf m}({\bf k}) + \frac{e {\bf \Omega}({\bf k})}{\hbar} \left[ \mu - \epsilon({\bf k}) \right] \right\} 
\end{equation}
where $t$ is the thickness of the monolayer MoS$_2$, ${\bf m}({\bf k})$ is the orbital magnetic moment, $\mu$ is the chemical potential and $\epsilon({\bf k})$ is the energy dispersion of the conduction band. Here we assume that MoS$_2$ is weakly $n$-doped with partially occupied conduction band. Weak spin splitting of the conduction band is ignored. When an external electric field ${\bf E}$ is applied, $f({\bf k})$ is given by $f({\bf k}) \approx f^{\rm eq}(\epsilon({\bf k})) + \partial_{\bf k} f^{\rm eq}( \epsilon({\bf k})) \cdot \frac{e}{\hbar} {\bf E}\tau$, where $f^{\rm eq} (\epsilon)$ is the equilibrium Fermi-Dirac distribution function and $\tau$ is the electron mean free time. Since  ${\bf M}$ vanishes in equilibrium with 
$f({\bf k}) = f^{\rm eq}(\epsilon({\bf k}))$, ${\bf M}$ in the presence of the driving field ${\bf E}$ becomes 
${\bf M} = \frac{1}{t} \frac{e \tau}{\hbar} E_a \left[ \int \frac{d {\bf k}}{(2\pi)^2} \, \partial_{k_a} f^{\rm eq}(\epsilon({\bf k})) \, {\bf m}({\bf k}) \right]$. Here it is used that $\partial_{k_a} f^{\rm eq}(\epsilon({\bf k}))$ is proportional to $\delta(\epsilon({\bf k})-\mu)$. Considering the relation ${\bf m}({\bf k})=\frac{e}{2 \hbar} \Delta^\prime {\bf \Omega}({\bf k}) \, || \, \hat{z}$ near the bottom of the MoS$_2$ conduction band, where $\Delta^\prime$ is the energy gap at $K$ and $K^\prime$ points, and by using the definition of the Berry curvature dipole, we obtain the relation
\begin{equation}
{\bf M} \, = \, -\frac{e^2 \tau}{2 t \hbar^2} \Delta^\prime ( {\bf D} \cdot {\bf E} ) \, \hat{z}.
\label{Eq2}\end{equation}
In this simplified equation, the valley magnetization  ${\bf M}$ is generated as a linear response to ${\bf E}$, which is regulated by the Berry curvature dipole ${\bf D}$. Note that the Berry curvature dipole was originally introduced in the quantum nonlinear Hall effect \cite{Sodemann2015,Low2015} which relates the Hall current density ${\bf J}_H$ as a quadratic response of $\bf{E}$: ${\bf J}_H \propto ({\bf D}\cdot{\bf E}) \hat{z} \times {\bf E}$. Interestingly, both ${\bf M}$ and ${\bf J}_H$ support that $({\bf D} \cdot {\bf E}) \hat{z}$ can be interpreted as an effective magnetic field. Out of the two ${\bf E}$'s responsible for ${\bf J}_H$, one ${\bf E}$ induces magnetization and effectively breaks the time-reversal symmetry and the other ${\bf E}$ induces the anomalous Hall effect in the presence of the magnetization.

\begin{figure}[t]
    \centering
    \includegraphics[width=0.5\textwidth]{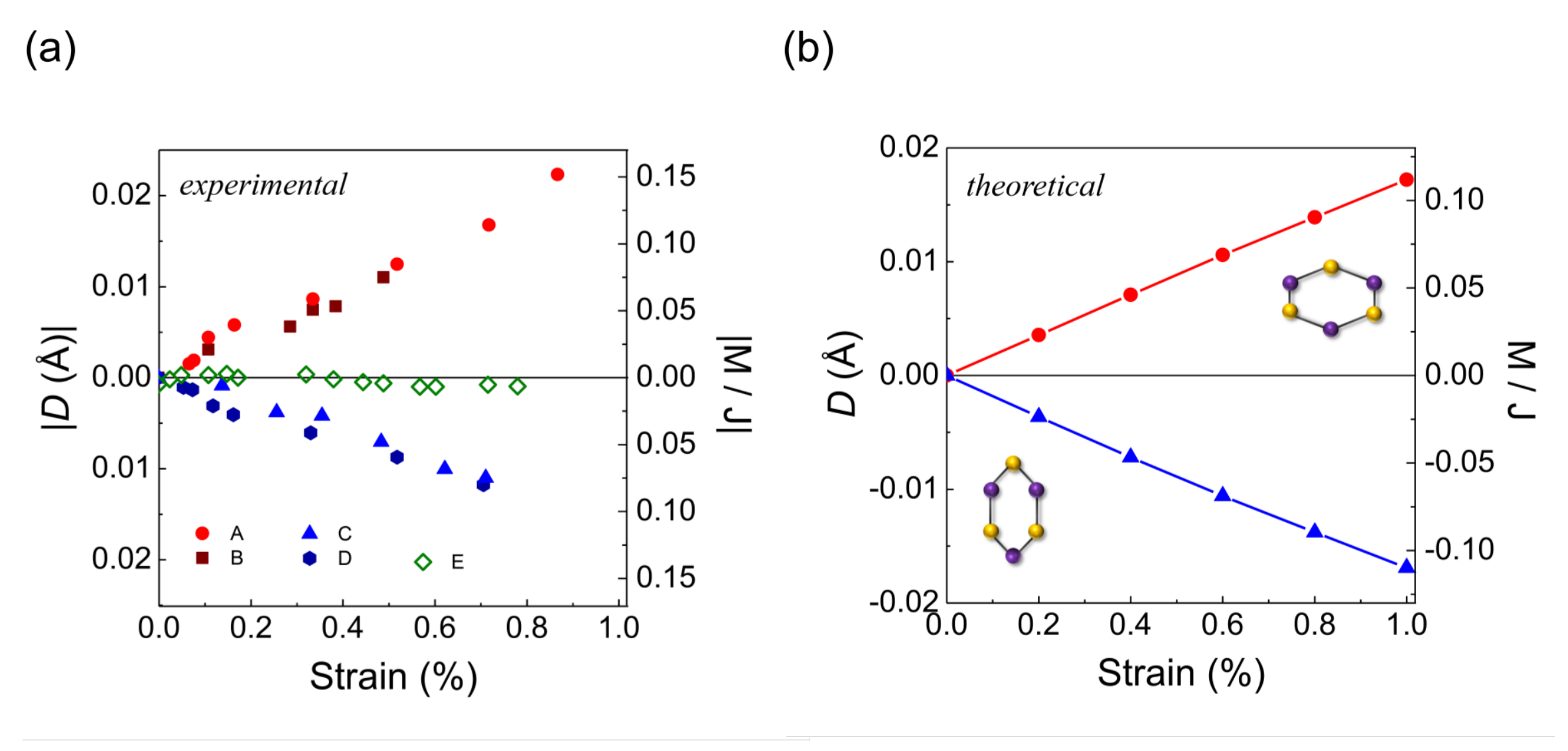}
    \caption{Experimental and theoretical Berry curvature dipole (${\bf D}$) and normalized valley magnetization (${\bf M}/{\bf J}$). {\bf (a)} $|{\bf D}|$ and $|{\bf M}/{\bf J}|$ measured from 5 different devices. Device A and B are strained along $\hat{x}_{\rm zigzag}$ while ${\bf E}$ is applied along $\hat{x}_{\rm zigzag}$. Device C and D are strained along $\hat{y}_{\rm armchair}$ while ${\bf E}$ is applied along $\hat{x}_{\rm zigzag}$. Device E is strained along $\hat{x}_{\rm zigzag}$ while ${\bf E}$ is applied along $\hat{y}_{\rm armchair}$. {\bf (b)}  Calculated ${\bf D}$ and ${\bf M}/{\bf J}$ when strain is applied along $\hat{x}_{\rm zigzag}$ (red circles) and $\hat{y}_{\rm armchair}$ (blue triangles). 
    }
    \label{Fig4}
\end{figure}

Eq. (\ref{Eq2}) explains many important features observed in our experiment. First, reversing the direction of ${\bf D}$ or ${\bf E}$ flips the direction of the out-of-plane magnetization ${\bf M}$. In Fig. \ref{Fig3}(e) and (f), we illustrate the band structures for two different directions of ${\bf D}$ under the same ${\bf E}$. The resulting ${\bf M}$ flips the sign upon reversing ${\bf D}$, which is clearly demonstrated in Fig. \ref{Fig3}(b) and (c). We have also checked the case when ${\bf E}$ is reversed while maintaining ${\bf D}$, which also changes the direction of ${\bf M}$ (Supplementary Materials Fig. S5). Second, nonzero ${\bf M}$ is generated only when  ${\bf D}$  and ${\bf E}$ have a collinear component. Since ${\bf D}$ is always generated along the zigzag direction of the crystal, a finite ${\bf M}$ requires ${\bf E}$ to be applied along the same zigzag direction. On the other hand, if ${\bf E}$ is applied perpendicular to ${\bf D}$, {\it i.e.}, along an armchair, no magnetization will arise even under a finite strain and an in-plane electric field. Such case is investigated with a device with the channel current flowing along the armchair direction of a crystal, from which no magnetization is measured with increasing strain (green diamonds in Fig. \ref{Fig4}(a)). Detailed device structure and measurement results can be found in Supplementary Materials Fig. S6.

The measured valley magnetization is also in a quantitative agreement with theoretical calculations. Fig. \ref{Fig4}(a) shows the magnetization normalized by current, $M/J$, which is converted from the measured value of $\theta_{\rm KR}$/$J$ in several different monolayer MoS$_2$ from the same bulk. In the conversion, we considered that $\theta_{\rm KR}$ is the result of the absorbance difference between right- and left-handed light at the probe wavelength. Such circular dichroism ($\rho$) can also be directly related to the Berry curvature dipole through $\rho = \frac{-\Delta^\prime e \tau }{\hbar {\it \Gamma}} {\bf E} \cdot  \int \frac{d {\bf k}}{(2\pi)^2} \, \partial_{k} f^{\rm eq}({\bf k}) \, {\bf \Omega}({\bf k})$, where ${\it \Gamma}$ is the thermal transition linewidth broadening \cite{Xiao2012,Lee2017,Yao2008} (Supplementary Materials Sec. 3.2). Note that the expression for $\rho$ is similar to that of the valley magnetization because both quantities have the same origin, {\it i.e.}, electron occupation enclosing asymmetric Berry curvature distribution at $K$ and $K^\prime$ valleys. The experimental $D$ is then obtained from $M/J$ in Eq. (\ref{Eq2}).

Fig. \ref{Fig4}(b) shows the calculated Berry curvature dipole using the tight-binding model when the Fermi energy is 30 meV above the conduction band minimum, roughly matching the doping density of our MoS$_2$ crystal, $n \sim 10^{13} \, {\rm cm}^{-2}$. The calculated $D$ and $M/J$ is linearly proportional to the magnitude of strain and has opposite signs for two different strain directions. The magnitude of theoretical Berry curvature dipole and normalized magnetization are in remarkable agreement with experimental results, suggesting that our method will also be applicable to probe the Berry curvature dipole in other materials.

\begin{figure}[t]
    \centering
    \includegraphics[width=0.5\textwidth]{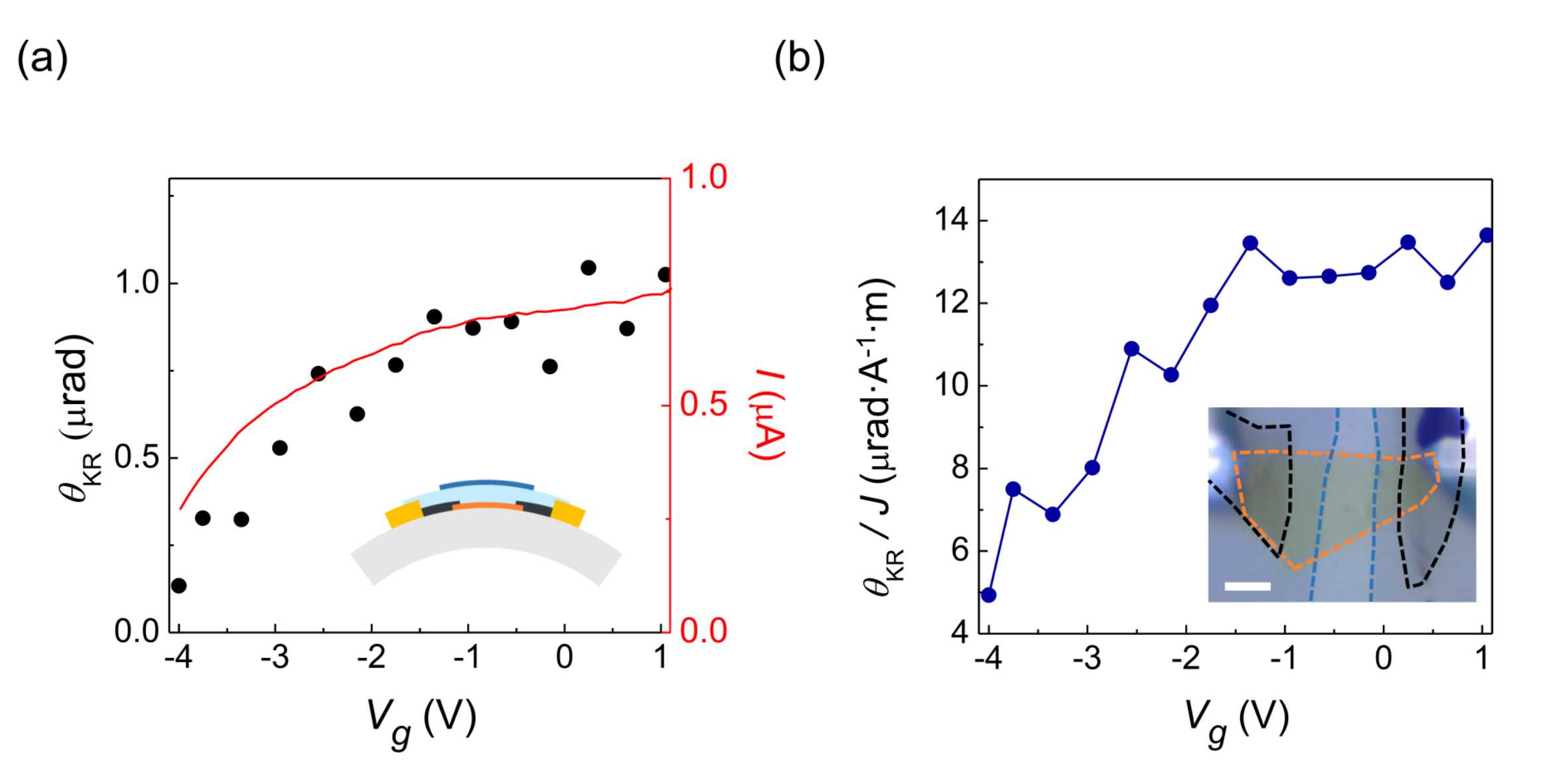}
    \caption{Valley magnetization generated in a flexible MoS$_2$ transistor. {\bf (a)} $\theta_{\rm KR}$ (black circles) and channel current (red line) measured as a function of the gate voltage at $V_{ds} = 1$ V under 0.28 \% strain. Inset: Strained device schematics with a graphene top gate (dark blue). MoS$_2$ (orange); {\it h}-BN (light blue); graphene source/drain (black); Cr/Au (yellow); PMMA (gray) {\bf (b)} $\theta_{\rm KR}/J$ as a function of the gate voltage at $V_{ds} = 1$ V under 0.28 \% strain. Inset: Microscope image of the fabricated transistor. MoS$_2$ (orange dashed line); graphene source and drain electrodes (black dashed line); graphene top gate (blue dashed line). Scale bar, 5 $\mu {\rm m}$.
    }
    \label{Fig5}
\end{figure}

Finally, we investigated the doping density dependence of the valley magnetization by fabricating a monolayer MoS$_2$ transistor. In this experiment, a FL graphene is additionally stacked on top of the MoS$_2$/graphene/{\it h}-BN heterostructure to apply a gate voltage as shown in Fig. \ref{Fig5}. A relatively thick {\it h}-BN ($\sim 80$ nm) is used as an insulating layer to avoid coupling between source and gate electrodes. Strain and in-plane electric fields in this device are applied in the same zigzag direction to produce substantial valley magnetization. Detailed strain dependent measurements are shown in the Supplementary Materials Fig. S7. Here we focus on the gate dependence of $\theta_{\rm KR}$ at a fixed strain of 0.28 \% when $V_{ds}=1$ V (Fig. \ref{Fig5}(a)). From the measurement, both $\theta_{\rm KR}$ and channel current increase with increasing gate voltage. In Fig. \ref{Fig5}(b), $\theta_{\rm KR}/J$ increases with gate voltage and eventually saturates. Our observation is well explained by the calculated valley magnetization as a function of Fermi energy (Supplementary Materials Fig. S11). Since the Berry curvature dipole represents the degree of asymmetry of Berry curvature distribution and because of the parabolic shape of band structure, the increasing rate of Berry curvature dipole with doping is the largest at the band edge.

In conclusion, our work shows that applying strain to monolayer TMD induces the Berry curvature dipole depending on the direction and magnitude of strain. The Berry curvature dipole enables the mechanical tuning of valley magnetization excited by an in-plane electric field. By employing strain as a new functionality, monolayer TMDs could potentially be useful for novel valley flexomagnetic and memory devices. Through advance in strain engineering of monolayer materials in combination with conventional technologies, {\it e.g.}, piezoelectric modulation \cite{Wu2014,Akinwande2014}, new opportunities for strain-mediated valley-charge conversion may also emerge. Our work further envisages the Berry curvature dipole as a tunable source of magnetization in other systems with quantum geometric properties \cite{Xu2018,Ma2018,Kang2019}.

The authors thank fruitful discussions with Kin Fai Mak and Jie Shan. The authors also acknowledge support from the National Research Foundation (NRF) of Korea (Grants No. 2017R1C1B2002631, No. 2018R1A5A6075964 and No. 2018R1A5A6086814). J. L. was supported by TJ Park Science Fellowship of POSCO TJ Park Foundation. Theoretical analysis of K.-H. K. and H.-W. L. was supported by the SSTF (BA-1501-07).

\end{document}